\pdfoutput=1
\documentclass[aps,twocolumn,prl,amsmath,amssymb,notitlepage,superscriptaddress,longbibliography,nofootinbib,longbibliography]{revtex4-1}
\usepackage[T1]{fontenc}

\usepackage{amsmath}
\usepackage{amssymb}
\usepackage{bm}
\usepackage{graphicx}
\usepackage{xcolor}
\definecolor{darkblue}{rgb}{0,0,0.6}
\definecolor{darkred}{rgb}{0.6,0,0}
\usepackage[%
  colorlinks=true,
  urlcolor=darkblue,
  linkcolor=darkred,
  citecolor=darkblue
]{hyperref}
\usepackage{url}

\begin{document}
\title{
Active-hydraulic flows solve the 6-vertex model (and vice versa) 
}

\author{Camille Jorge}
\email{camille.jorge@ens-lyon.fr}
\affiliation{Univ. Lyon, ENS de Lyon, Univ. Claude Bernard, CNRS, Laboratoire de Physique, F-69342, Lyon.}
\author{Denis Bartolo}
\email{denis.bartolo@ens-lyon.fr} 
\affiliation{Univ. Lyon, ENS de Lyon, Univ. Claude Bernard, CNRS, Laboratoire de Physique, F-69342, Lyon.}

\begin{abstract}
By confining colloidal active fluids in microchannel networks, we demonstrate that their degenerate flows corresponds to the configurations of the six-vertex model. 
We use this quantitative correspondence to control and explain the active flows that emerge in square grid networks. 
In particular, we show that the Lagrangian trajectories of active particles realize the Baxter-Kelland-Wu mapping and form completely packed loops, whose geometry can be exactly predicted and explained. 
We then go beyond the square-grid geometry and introduce a general framework for predicting the geometry of active-hydraulic flows in arbitrary networks.
\end{abstract}

\maketitle

\begin{figure*}
    \includegraphics[width=\textwidth]{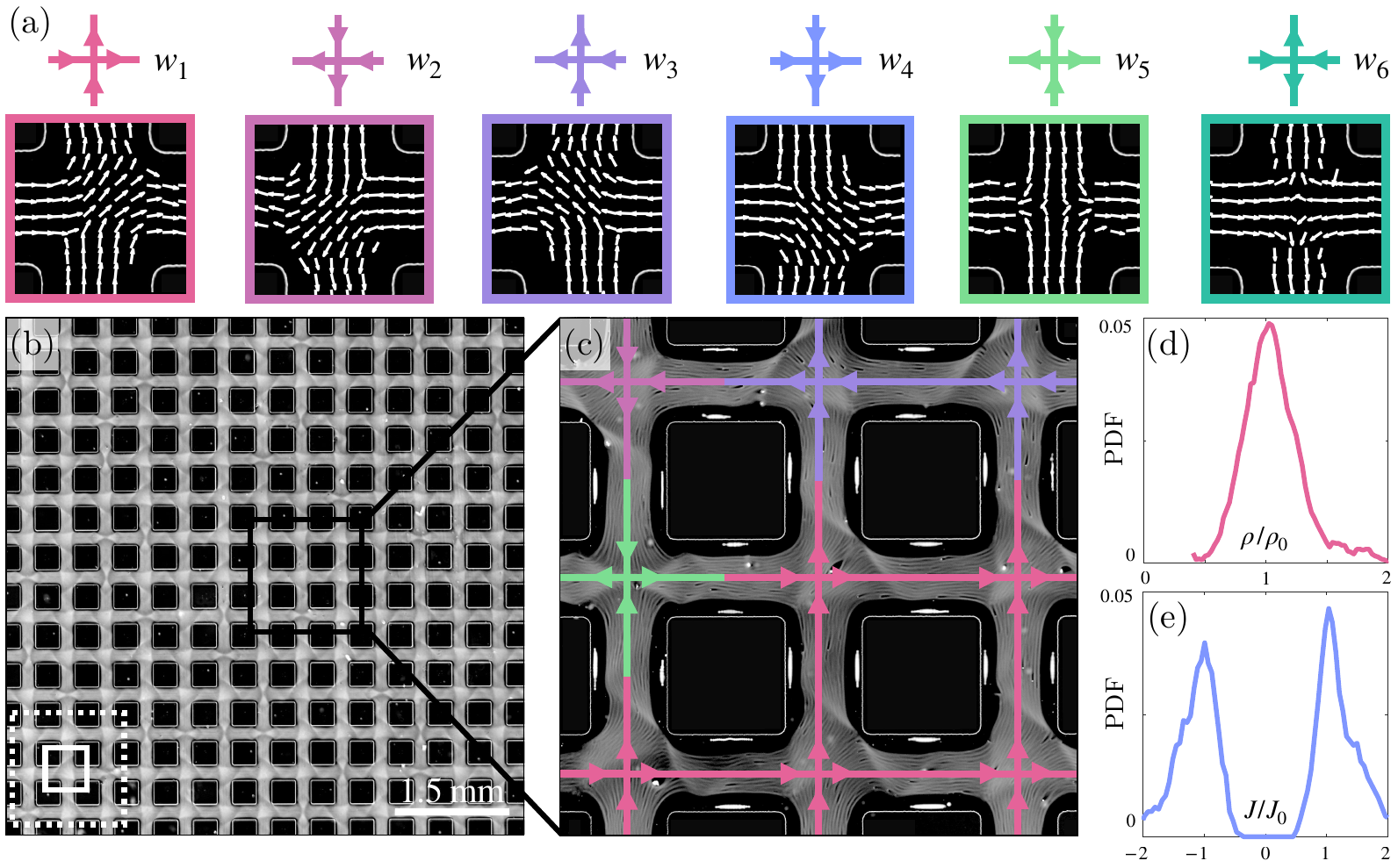}
        \centering
    \caption{
    {\bf The six-vertex-model rules are realized by active hydraulics flows. }
    {\bf (a)} Top row: Illustrations of the  vertices of the six-vertex models. The $w_i$ are their statistical weight.
    Bottom row: Velocity field at the nodes of the channel network. The   six possible flow configurations  are in a one-to-one correspondence with the six-vertex model. 
    {\bf (b)} Active fluid assembled from colloidal rollers confined in a square grid of identical channels. The  white squares illustrate the relative sizes of the whole grid (dashed line) and of the picture of the $13 \times 13$ channels (solid line). The active flow is stationary. 
    {\bf (c)} A superimposition of subsequent images show the trajectories of the active colloids. The arrows indicate the direction of the flow, and the colors indicate the type of vertices, see {\bf (a)}.
    {\bf (d)} Probability density (PDF) of the colloid density, $\rho$ measured in all the channels of the network. 
    The number density is normalized by its average values $\rho_0=8500\, \rm mm^{-2}$.
    {\bf (e)} Probability density of the current, $J$. The current value is normalized by its average value $J_0=1300\pm 260\, \rm s^{-1}$.
     } 
    \label{Fig1}
\end{figure*}

From the 2D Ising model to quantum spin chain, exactly solvable models  are rarely intended to provide accurate descriptions of real-life experiments. 
One of their essential values is to reveal profound connections between seemingly unrelated areas of physics and mathematics. 
The six-vertex model offers a compelling example~\cite{Lieb1967,Baxter2016}.
This model was inspired by Pauling's early work, and intended to gain some insight into the residual entropy of water ice~\cite{Pauling35}. 
It consists in assigning different statistical weight to the six vertices drawn in Fig.\ref{Fig1}a, and to place them on the square lattice. 
Although it cannot explain actual ice-calorimetry measurements, the six-vertex model has become a cornerstone of mathematical physics.   
It has indeed revealed strong links between fields as diverse as Coulomb gas physics, percolation theory, ferroelectricity, polymer physics, loop models, quantum hall physics, conformal field theory, combinatoric and stochastic processes, see e.g.~\cite{Zuber88,Jacobsen2009,zinn2009six,jacobsen1998field,duplantier2000conformally,kuperberg1996another}.

Here, we show that the six vertex model provides an accurate description of active matter experiments. Confining colloidal active fluids in  hydraulic networks, we  establish experimentally a quantitative correspondence between the active flow field and the spin configurations of the six-vertex model.
We  then use this one-to-one correspondence to quantitatively explain the seemingly random geometry of active hydraulic flows in square-grid networks. 
In particular, we show  that the Lagrangian trajectories  follow Baxter-Kelland-Wu loops and correctly predict their self-similar morphologies~\cite{baxter1976equivalence}. 
Informed by our findings, we  conclude with the presentation of a comprehensive set of active-hydraulic laws  that  should predict the geometry of active flows in arbitrarily complex networks.

In our  experiments, we use  2D active fluids assembled from Quincke rollers of diameter $5\,\rm \mu m$~\cite{Bricard2013}. 
As illustrated in Fig.~\ref{Fig1}b, we first confine them in a square-grid network made of 1,800 straight channels (width $200\,\rm\mu m$), see~\cite{Jorge2024} for a detailed presentation of the experimental methods. 
As we turn on the electric motors that powers the self-propulsion of our Quincke rollers, they self-assemble into an active liquid  whose  spontaneous flows quickly reach a steady state illustrated in Figs.~\ref{Fig1}b and \ref{Fig1}c and Supplementary videos 1 and 2.

The steady flows are homogeneous. 
We find that the number density is narrowly peaked on $\rho_0= 8,500\,\rm  mm^{-2}$, and that the particle flux $J$ hardly fluctuates around two possible values $J=\pm J_0$, where $J_0=1,300\,\rm s^{-1}$, see Figs.~\ref{Fig1}d and~\ref{Fig1}e and~\cite{Morin2018}.
The resulting flow patterns can therefore be accurately described by using the sign of $J$ to orient the edges of the graph formed by the  channel network, see Fig.~\ref{Fig1}e and \cite{Woodhouse2016stochastic}. 
Repeating the same experiment, with the same colloids, in the same geometry, however  yields different patterns all having a seemingly random geometry. 
In this letter, our goal is to explain the morphology of these emergent flows.

We can make an important progress in this direction by noting that the geometry of the current field at the nodes  exactly correspond to the six vertices of the six-vertex model, Fig.~\ref{Fig1}a.
This observation is a direct consequence of the two constraints imposed to all confined active flows. 
They were first identified by Woodhouse and Dunkel in~\cite{Woodhouse2016stochastic}, and confirmed in a number of experiments: 
(i) When confined in sufficiently slender channels the flows of active fluids are laminar and operate at constant speed set by the level of activity of the materials~\cite{WiolandNJP2016,Wu2017,Morin2018,Hardouin2020}. (ii) Mass is conserved. Therefore, in steady state, when the density heterogeneities have relaxed, activity and mass conservation impose
\begin{align}
    J_{\langle i,j\rangle}&=\pm J_0,
    \label{eq:hydro1}
    \\
    \sum_jJ_{\langle i,j\rangle}&=0
    \label{eq:hydro2}
\end{align}
at  every edge ${\langle i,j\rangle}$ and node $i$. 
In other words, in all networks with a fourfold coordination, irrespective of the channel and node geometry, all vertices must obey the two-in-two-out rule illustrated in Fig.~\ref{Fig1}a.
Eqs.~\eqref{eq:hydro1} and~\eqref{eq:hydro2} are however not sufficient to explain our active hydraulic flows. 
They define the vertices of a vertex model \cite{Baxter2016,Udagawa2021}, but not their statistical weight. 
Early theories  assumed that they had equal probabilities~\cite{Woodhouse2016stochastic,Woodhouse2017, Woodhouse2018information}, but
our measurements do not confirm this spin-ice hypothesis. 
After repeating ten times the same experiment, we count the occurrences of each vertex and measure their statistical weight $w_i$, $i=1,\ldots,6$.
We find that they are not identical: $w_{1}=w_{2}=w_{3}=w_{4}=0.20\pm 0.01$ while $w_{5}=w_{6}=0.09\pm 0.01$. 
These weights are not only material dependent, but also strongly depend on the geometry of the junctions between the straight channels \cite{Wioland2016}. 
\begin{figure*}
    \includegraphics[width=\textwidth]{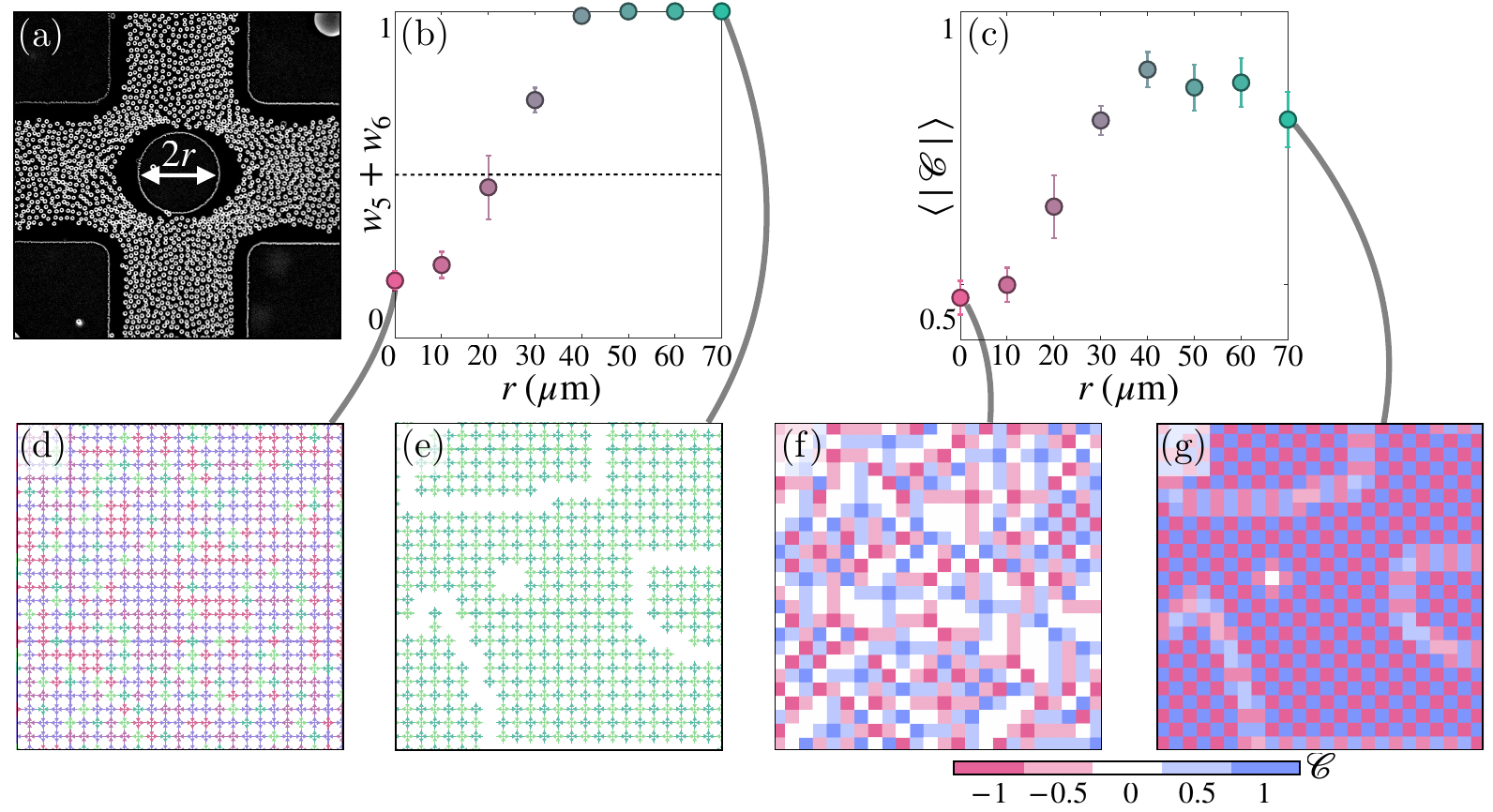}
        \centering
    \caption{{\bf Weight of the six vertices and order-to-disorder transition. }
        {\bf (a)} Image of a junction with a circular post of radius $r=70\,\rm\mu m$.
    {\bf (b)} Evolution of $w_5+w_6$ with the post radius $r$. The dashed line corresponds to $w_5+w_6=\frac{1}{2}$.   %
    {\bf (c)} Evolution of the average circulation $\mathcal C=\langle| C_i|\rangle_i $, where  $C_i$ is the circulation around the i$^th$ plaquette. 
     {\bf (d)} Example of a flow field measured in an experiment where $r=0\,\rm\mu m$.
     {\bf (e)} Example of a flow field measured in an experiment where $r=70\,\rm\mu m$. 
     When $r$ is large, the flow only includes vertices of type 5 and 6.
     {\bf (f)} Circulation field measured in an experiment where $r=0\,\rm\mu m$ (disordered).
     {\bf (g)} Circulation field measured in an experiment where $r=70\,\rm\mu m$ (ferro-electric phase).} 
\label{Fig2}
\end{figure*}
To see this, we add cylindrical posts at each junction (Fig.~\ref{Fig2}a) and measure the evolution of the $w_i$s with the post radius $r$ (Fig.~\ref{Fig2}b). 
For all  radii,  we have $w_1=w_2=w_3=w_4$ and $w_5=w_6$ as expected from the axisymmetry of the post shape. 
But Fig.~\ref{Fig2}b shows that $w_5+w_6$ monotonically increases from $\simeq 0.18 \pm 0.03$ to $1$ as $r$ increases.
From a fluid mechanics perspective, this results is intuitive: adding a solid obstacle at a junction favors the formation of a stagnation point.
The variations of the $w_i$s are however not anecdotal. 
They are accompanied by a qualitative evolution of the macroscopic flow patterns as shown in Figs.~\ref{Fig2}d and \ref{Fig2}e. 
To explain the geometries of the spontaneous flows, we make a second key observation. 
Eqs.~\eqref{eq:hydro1},~\eqref{eq:hydro2} and the $\lbrace w_i\rbrace$ fully specify the statistics of the flow field: our active hydraulic experiment generate independent configurations of the six vertex model. 
We can therefore use the large body of knowledge on this integrable model to  quantitatively explain the morphology our hydraulic patterns.
A first well established result concerns the phase behavior of the six-vertex-model. 
When $(w_5+w_6)=\frac1 2$ theory predicts an order-to-disorder transition between an antiferroelectric and an isotropic phase. 
We do observe this structural transition. 
Antiferroelectric order corresponds to edge fluxes of alternate signs, or equivalently to plaquettes of alternate circulations, as in a cellular flow \cite{young2007stretch, wandersman2010buckled}. 
In Fig.~\ref{Fig2}c we plot the evolution of the average circulation $\mathcal C$ as a function of the post radius $r$. We show the circulation field for $r =0 \, \rm \mu m$ and $r =70 \, \rm \mu m$ in Figs.~\ref{Fig2}f and \ref{Fig2}g.
We find a clear sign of a transition  when $w_{5,6}(r)=\frac 1 2w_{1-4}(r)$ (see also Fig. 2b).
This phase transition has clear consequences on the geometry of the hydraulic flows and, in particular, on their Lagrangian trajectories illustrated in Fig.~\ref{Fig3}a.
 
\begin{figure*}
    \includegraphics[width=\textwidth]{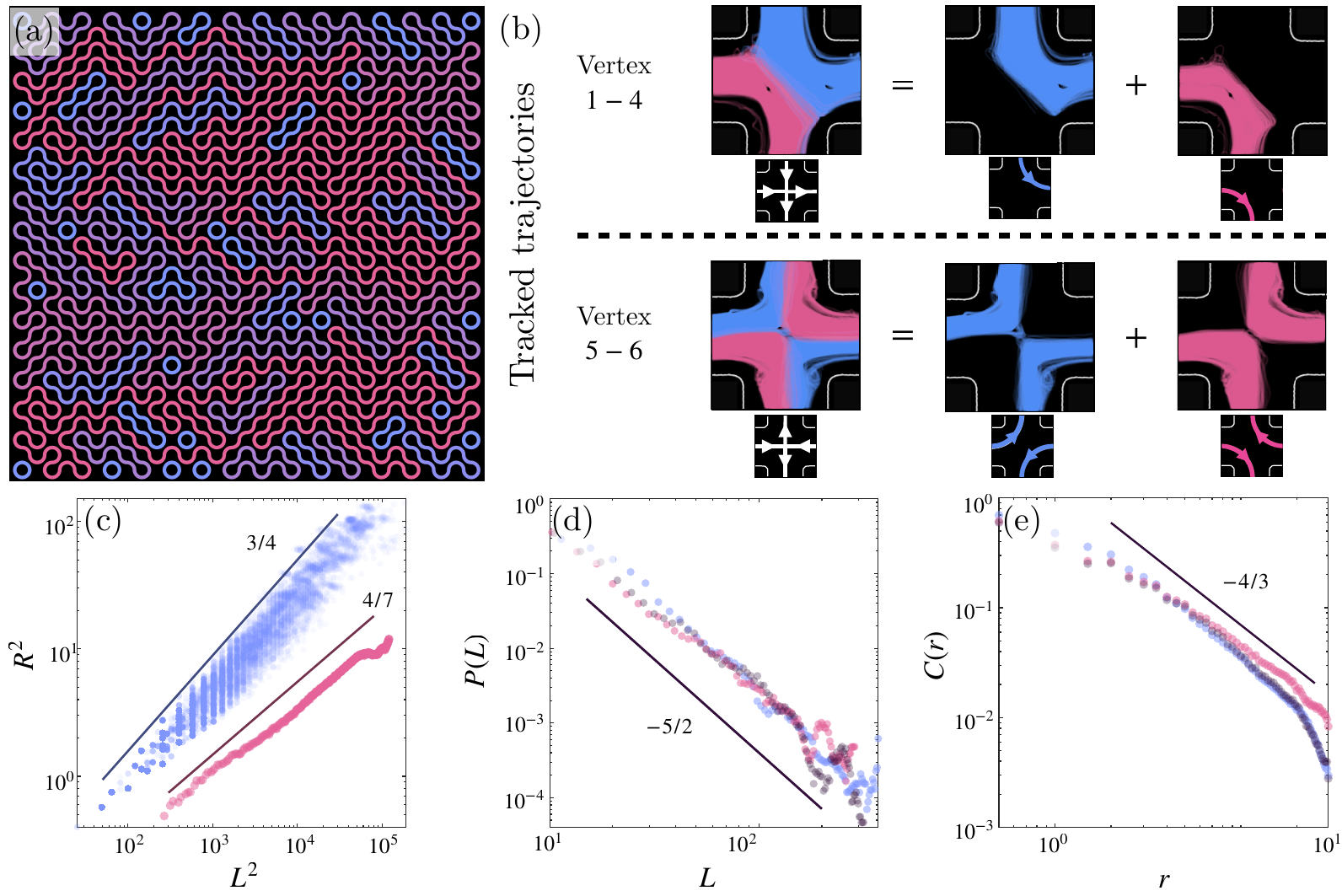}
        \centering
    \caption{{\bf Lagragian trajectories, self-avoiding loops and  geometry of the disordered flows.  }
        {\bf (a)} Experimental Lagrangian trajectories of the active flow in the disordered regime (no circular post). 
        The trajectories are constructed from the experimental flow field using the   rules obeyed by the rollers at every nodes, see {\bf (b)}.
    {\bf (b)} Tracked trajectories of the colloidal rollers. 
    Labeling the trajectories with two colors, we find a direct correspondence with the Baxter-Keeland-Wu mapping from the six-vertex-model configurations to closely packed loops on the square lattice~\cite{baxter1976equivalence}. 
    $93\%$ of the 155,000 trajectories we have tracked follow these rules.
     {\bf (c)} Symbols: Gyration radius $R_{\rm g}$ of the Lagrangian trajectories plotted versus their length $L$.  Blue symbols: All loops but the largest. Solid lines: theoretical predictions of the $\nu$ exponent.
     Pink symbols: largest loop measured in each experiments. For a better statistics, for the largest loop, we plot $R^2(\ell^2)=\langle |\mathbf r(s)-\mathbf r(s+\ell)|^2\rangle_s$, were $\mathbf r(s)$ is the position measured at the curvilinear coordinate $s$ along the loop.
     Different experiments correspond to symbols having a different opacity. 
     {\bf (d)} Probability $P(L)$ of finding a loop of length $L$ in the ensemble formed by the Lagrangian trajectories.
     Symbols: experiments. Solid line: Theoretical prediction. 
       {\bf (e)} Two point correlation function $C(r)$. It corresponds to the probability of finding two points separated by a distance $r$ within the same Lagrangian loop. Symbols: experiments. Solid line: Theoretical predictions.
       {\bf(d)} and {\bf (e)}
          The three colors correspond to  three different post radii ($r=0,\, 5,\,10\,\rm \mu m$). The statistics corresponds to ten independent realization of the experiment. The collapse of our data indicate that all the disordered flows belong to the same universality class.
    } 
\label{Fig3}
\end{figure*}
In the ordered state, the active fluid circulates around the plaquettes and only two stationary flows can exist (away from possible domain walls). 
The resulting Lagrangian trajectories obviously reduce to closed loops of length 4, see Supplementary video 3.
The disordered phase is mode subtle: it corresponds to a critical phase where seemingly random fluctuations correlates over system-spanning scales. 
To better understand its geometry, we track the active colloids and plot their trajectories at the two type of vertices in  Fig.~\ref{Fig3}b. 
We find that the Quincke rollers propel along trajectories that realize the so-called Baxter-Kelland-Wu mapping from the 6-vertex to the Completely Packed Loop model (CPL) \cite{baxter1976equivalence}. 
This observation readily tells us that these Lagrangian trajectories  should form closely packed soups of self avoiding loops, in good agreement with our observations (Fig.~\ref{Fig3}a).
Predicting the geometry of the emergent flows therefore amounts to determining the statistics of the loop geometry. 
We explain in SI how to quantitatively describe these geometries thanks to a series of mappings between the six-vertex, CPL and bond percolation models for which a host of exact results have been derived \cite{baxter1976equivalence}. 
These theories predict that the largest loop has the same statistics as the hull of the percolation cluster at criticality \cite{saleur1987exact}. 
It has a self-similar shape and its gyration radius scales as $R_{\rm g}\sim L^\nu$ with $\nu=4/7$ (Fig.~\ref{Fig3}c). 
All the other loops are less crumpled and their gyration radius is that of self-avoiding polymers with $\nu=3/4$ \cite{saleur1987exact, kolb1993loop, aizenman1999path}. 
We also find that the probability of finding a Lagrangian trajectory of length $L$ should scale as $P(L)\sim L^{-5/2}$ \cite{saleur1987exact, kondev1996operator} (Fig.~\ref{Fig3}d). 
Lastly, the two point correlation function $C(r)$ that corresponds to the probability of finding two point at a distance $r$ within the same loop should scale as $C(r)\sim r^{-4/3}$ \cite{saleur1987exact, kondev1996operator}  (Fig.~\ref{Fig3}e).
Figure~\ref{Fig3}c compares these analytical predictions to our experimental measurements. The agreement is excellent.
We therefore conclude that the  vertex rules of Fig.~\ref{Fig1}a define the laws of active hydraulics in tetravalent pipe networks.
From a dual perspective, our experiments show that the Quincke rollers collectively solve the six vertex model. 

We close our letter with a more general discussion.
Solving hydrodynamic equations, or particle based models would limit the predictive power of  simulations to hydraulic networks composed of a handful of channels.
We propose to solve this computational problem by introducing a comprehensive set of active hydraulics rules, beyond the specific of the square-grid network. 
We construct these rules from the results presented above and in Ref.~\cite{Jorge2024}, they apply  to networks made of channels of identical cross sections but having arbitrarily complex geometries.
(i) The first rule is given by mass conservation, Eq.~\eqref{eq:hydro1}.
(ii) The second rule is the active-flow law and generalizes Eq.~\eqref{eq:hydro1}. 
In general, the local current can take three different values $J_{\langle i,j\rangle}=-J_0\,,0,\,+J_0$. 
When $J=0$, the active fluid flows at the sub channel scale but forms vortices and do not support any net flux~\cite{Jorge2024}.
(i) implies that at least one of the edge currents  vanishes at every node having an odd coordination number: laminar flows are geometrically frustrated~\cite{Woodhouse2017,Jorge2024}.
We note that this frustration can also have a dynamical origin as seen in Figure~\ref{Fig2}g where the domain walls between incompatible antiferroelectric regions host localized vortices\footnote{To predict the morphology and statistics of the domain walls, the vertex rules of Figure~\ref{Fig1}a should be complemented by 13 additional vertices including an even numbers of edges where $J=0$. This corresponds to the 19-vertex model that has also been extensively studied, see e.g.~\cite{knops199419}. }.
Rules (i) and (ii)  define vertex rules. 
However, as in all vertex problems, they  must be complemented by the statistical weights of the vertices (iii) and interactions rules  (iv). 
They are intrinsically determined by the microscopic geometry of the nodes and channels.
(iii)  Weight of the vertices: it is determined by the geometry of the nodes. They must be measured experimentally. 
A simple control strategy consists in adding  flow splitters to bias the flow statistics, 
and promote spacial order, see Figs.~\ref{Fig2}e and ~\ref{Fig2}g.  
(iv) Vertex interactions: they are determined by the channel geometry and must be measured experimentally.
As the fluid exits a vertex and enters a channel where $J\neq0$, the deformations of the active flow that formed at the vertex relax to yield a laminar flows. They  do not propagate to the adjacent vertex, see Fig. 1e. 
As a result  two vertices connected by an edge where $J\neq0$ are hardly coupled. 
This reasoning explains why the rules (i), (ii) and (iii) are enough to account for all our measurements in square-grid networks. 
However along edges where $J=0$ vortices forms.  The continuity of the flow field at the subchannel level hence  couples the handedness of the vortices to the flow configurations in the  adjacent nodes. This coupling defines the vertex interaction rules discussed in~\cite{Jorge2024}. 

With these four rules at hand, we can then generate the active-flow configurations using any algorithm that minimizes a cost function to enforce (i)--(iv),  
without having to deal with the computationally expensive resolution of the  flows at the sub-channel scale.

From a more fundamental perspective, active hydraulics is  in a one-to-one correspondence with the statistical mechanics of interacting loop models that is a topic of intense studies both in theoretical  physics and mathematics.
To date, however, most  predictions focus on 2D loop ensembles and therefore to 2D active hydraulic. 
 Predicting the geometry of the active flows in tridimensional pipe networks hence remains both a formidable theoretical and experimental challenge.\\

\begin{acknowledgments}
 We thank Am\'elie Chardac and Alexis Poncet for fruitful interactions. 
   This work was partly supported by the European Research Council (ERC) under the European Union’s Horizon 2020 research and innovation program (grant agreement No. [101019141]) (DB and CJ).
\end{acknowledgments}

\bibliographystyle{apsrev4-1}
\bibliography{biblio}

\end{document}


\title{
Active-hydraulic flows solve the 6-vertex model (and vice versa)\\ Supplementary Informations
}

\author{Camille Jorge}
\affiliation{Univ. Lyon, ENS de Lyon, Univ. Claude Bernard, CNRS, Laboratoire de Physique, F-69342, Lyon.}
\email{camille.jorge@ens-lyon.fr}
\author{Denis Bartolo}
\affiliation{Univ. Lyon, ENS de Lyon, Univ. Claude Bernard, CNRS, Laboratoire de Physique, F-69342, Lyon.}
\email{denis.bartolo@ens-lyon.fr}

\maketitle

\renewcommand{\theequation}{S\arabic{equation}}
\renewcommand{\thefigure}{S\arabic{figure}}

\section{Experiments}

\subsection{Measurement of the density and current fields}

\label{subsec:fluo}

We measure the density and current fields by adding  $0.6 \%$ of fluorescent PS spheres  of radius $5\mu \rm m$ (Reference).  They are identical to the non-fluorescent rollers, we can therefore use  them as flow tracers. Their excitation and emission wavelengths are $580 \, \rm nm$ and $605 \, \rm nm$. 

\begin{figure*}    
\includegraphics[width=\textwidth]{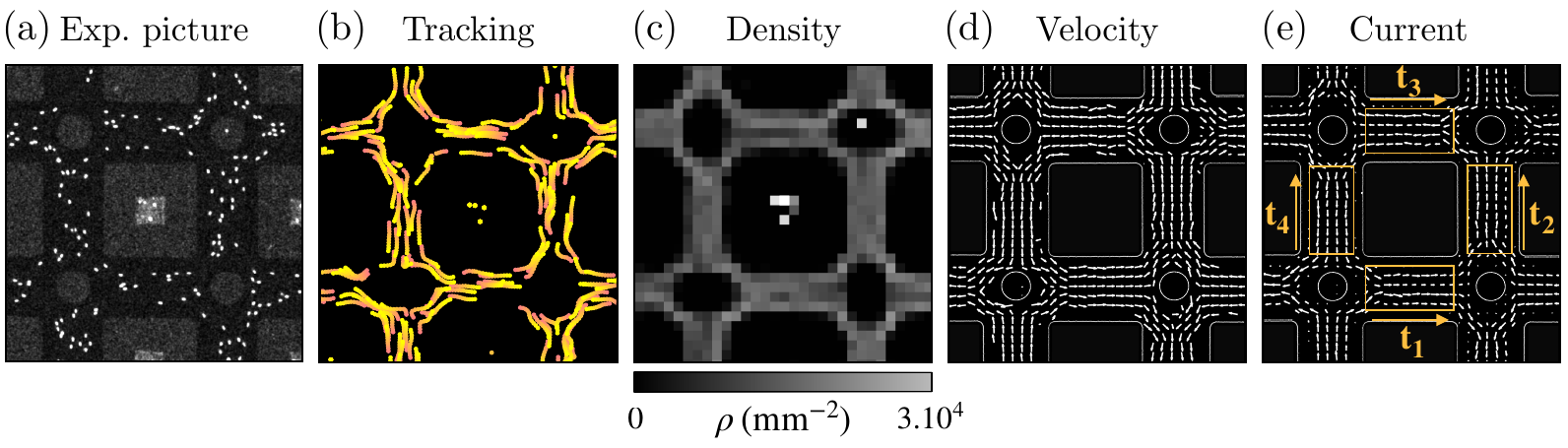}
    \caption{{\bf Measurements of the density, flow and current fields.}
    {\bf (a)} Snapshot of an active flow in found connected channels. The fluorescence  image shows only a small fraction of the colloidal rollers ($0.6\%$). 
    {\bf (b)} Tracked trajectories.
    {\bf (c)} Number density of the colloidal rollers defined as the overall number of particle in square boxes of size $64\,\rm \mu m$.
    {\bf (d)} Velocity field measured from PTV measurements on the tracked trajectories.
    (e) Current field reconstructed from the density and flow fields. The $J_i$s are scalar quantities defined as the average current in each rectangular windows. 
    Positive currents are associated to flows that point in the direction of the $t_i$ vectors showed on the image. 
    }
    \label{fluo}
\end{figure*}

We observe the fluorescent particles   using a Nikon AZ100 microscope with a magnification of $2.4$ and a 100 Watt full spectrum HBO lamp lamp (Nikon). We filter the fluorescence signal of the tracers using a dichroic mirror and a narrow band filter. 
We record the steady flow for $3 \, \rm s$ with a Hamamatsu ORCA-Quest qCMOS camera. 
We set the frame rate to $100 \, \rm fps$ and the observation window is a square with side length $4.6 \, \rm mm$, see Fig.~\ref{fluo}a. We detect the tracers using a detection algorithm inspired from \cite{lu2007target}, and we track them with the algorithm from ref. \cite{blair2008matlab}, see Fig.~\ref{fluo}b.

We measure the density field $\rho$ based on the position  of the particles. We count the number of particles in square boxes of size $64 \, \rm \mu m$ arranged on a square lattice (lattice spacing $32 \, \rm \mu m$).
We then average the density field  over $3 \, \rm s$, see Fig.~\ref{fluo}c. To measure the average density in each  channel by performing a local average.
We measure the velocity of the particles based on their displacements between two subsequent frames. 
We construct the Eulerian velocity field $\bf v$ by performing the same averaging operation as for the density field, first in boxes of $64 \, \rm \mu m$ spaced by $32 \, \rm \mu m$, and then temporally over a period of $3 \, \rm s$, see Fig.~\ref{fluo}d. 
The current field is $\rho \times {\bf v}$. We measure the average current per channel by spatially averaging $\rho \times {\bf v}$ within each channel, see Fig.~\ref{fluo}e.

\subsection{Construction of the streamlines}

\label{subsec:streamlines}
\begin{figure}[!h]
\includegraphics[width=\columnwidth]{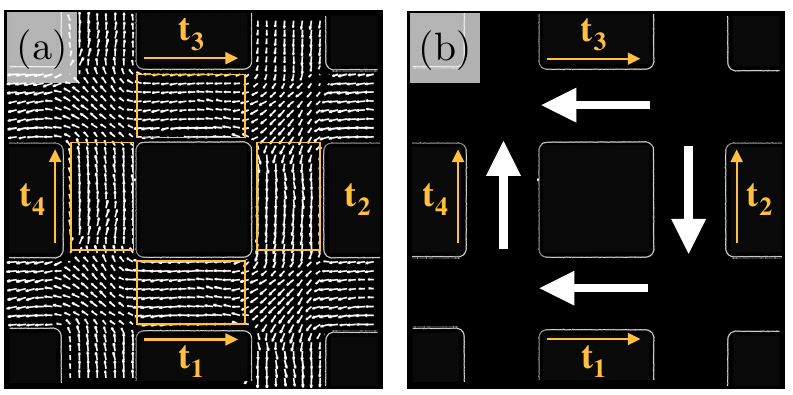}
    \caption{{\bf Current field from PIV measurements}
    {\bf (a)} PIV measurement of the time average of the colloidal roller flow in steady state.
    {\bf (b)} Current ($J$) averaged in space and time.}  
    \label{streamlines}
\end{figure}

To measure the velocity field in the whole network, we record bright field images of the flow over $3\,\rm s$ using a LUX160 (Ximea) camera mounted on a Nikon AZ100 microscope. The magnification is $2.4$.
The frame rate is $200 \, \rm fps$. We then measure the velocity field using the standard PIV algorithm PIVLAB (Matlab) \cite{stamhuis2014pivlab}.
The PIV-box size is $52 \times 52 \, \rm \mu m^2$, and the spacing between two boxes is set to half the box size.

Since the magnitude of the current in all the channels is narrowly peaked on (with $J_0= 1,300\,\pm 170\, \rm s^{-1}$), we represent the edge current  by a vector ${\bf J}_i$ of magnitude $1$ or $0$ (in the few channels where vortices form and suppress any form of net transport). 
In practice, we define ${\bf J}_i$ as follows. We measure the average velocity field direction $\langle {\bf v} \rangle_i$ in the channel $i$, see Fig.~\ref{streamlines}a. 
We define ${\bf t}_i$  the unit  vector tangent oriented along the centerline of the channel $i$ (Fig.~\ref{streamlines}b). 
If the magnitude of the scalar product $| \langle {\bf v} \rangle_i \cdot {\bf t}_i |$ exceeds $0.5$, we then set ${\bf J}_i = {\rm sign}(\langle {\bf v} \rangle_i \cdot {\bf t}_i) {\bf t}_i$, and ${\bf J}_i = 0$ otherwise, 

\subsection{Trajectories at junctions}

\label{subsec:jonctions}

\begin{figure}[!h]
    \includegraphics[width=\columnwidth]{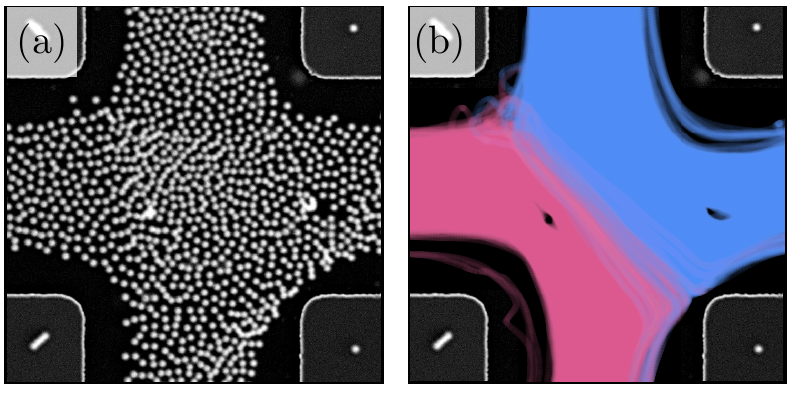}
    \caption{{\bf Trajectories at junctions.} 
    {\bf (a)} Snapshot of an active flow at junctions. The channel width is $200 \, \rm \mu m$.
    {\bf (b)} Lagrangian trajectories at the same junction. Pink : particles particles coming from the left channel. Blue : particles coming from the top channel. 
    }
    \label{SI_traj}
\end{figure}

It is technically impossible for us to track several hundred thousand particles with a diameter of $5 \, \rm \mu m$ in a network of $\sim 4 \, \rm cm^2$. However, we can obtain Lagrangian trajectories indirectly. Since in the channels the Quincke rollers move in straight lines, we only need to understand their trajectories at the junctions of the microfluidic network.
To achieve this, we zoom in on the flow at a junction. We use the same equipment as in subsection \ref{subsec:streamlines}, but this time with a magnification of $6$, which provides good resolution on the colloids (see Fig.\ref{SI_traj}a). We perform a $20 \, \rm s$ long acquisition at a frequency of $300 \, \rm fps$. We then detect and track the colloids. We can then measure the Lagrangian trajectories at the junctions. In Fig.\ref{SI_traj}b, we represent in pink the particles coming from the left channel and in blue those coming from the upper channel. We observe that these trajectories correspond in the vast majority of cases ($93\%$ for a sample of $155,000$ trajectories) to the Baxter-Kelland-Wu mapping.

\section{Data analysis}

\label{sec:data_analysis}





\subsection{Lagrangian trajectories from Eulerian measurements}

\subsubsection{Sampling}

\begin{figure*}
\includegraphics[width=\textwidth]{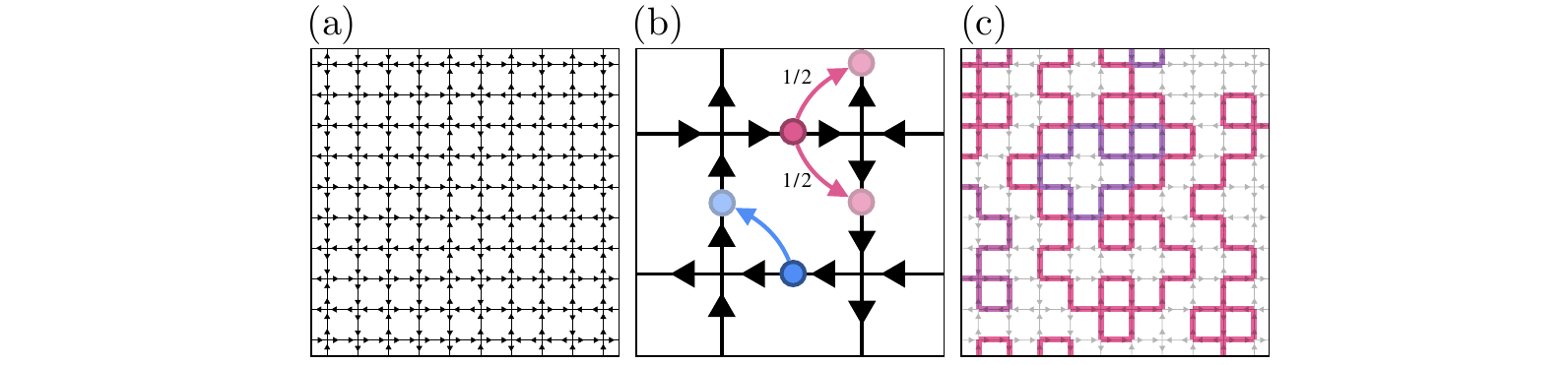}
    \caption{{\bf Construction of the Lagrangian trajectories.}
    {\bf (a)} Experimental current map ($r=0\,\rm \mu m$).
    {\bf (b)} Illustration of the algorithm used to construct the Lagrangian trajectories.
    {\bf (c)} Examples of eight Lagrangian trajectories, all forming closed loops 
    }
    \label{traj_euler}
\end{figure*}

We  define the Lagrangian trajectories associated to the flow field as follows. 
As shown in the main text, we can define Lagrangian trajectories from the spin  of the current field using the Baxter-Kelland-Wu cuts. 
This decomposition is however not bijective — a current map does not generate a unique set of Lagrangian trajectories— we therefore need to sample all possible Lagrangian paths. 
In practice, we sample the Lagrangian trajectories by implementing random walks on the current map (Fig.~\ref{traj_euler}a). The rules that define  the random walks are:
\begin{itemize}
    \item The accessible positions for the random walkers are the centers of the channels.
    \item The walkers follow the current vectors.
    \item At a vertex a walker can only make turn. If the junction is of type $5,6$, it  turns left or right with equal probabilities, see Fig.~\ref{traj_euler}b.
    \item Once the walker returns to its starting point, the walk ends.
\end{itemize}

In Fig.~\ref{traj_euler}c, we show some Lagrangian trajectories constructed with the above procedure, they correspond to the current map of Fig.~\ref{traj_euler}a. We stress that these loops are different from the actual trajectories followed by tracers that can jump from one streamline to another due to their orientational diffusion. At long times tracers explore multiple Lagrangian loops.

\subsubsection{Statistical Weighting of Loops}

The correspondence between the flow (6-vertex model) and the trajectories (loops) is not bijective due to the vertices of type $5$ and $6$. 
Our sampling protocol allows us to measure \textit{all} possible trajectories from a given flow configuration. 
Therefore, the edges of the network can be explored by multiple loops resulting in an oversampling of the largest loops.
To correct for this unwanted  bias we define the statistical weight $w$ of a loop as follows:

\begin{equation}
\displaystyle w = \frac{1}{L} \sum_{e} \frac{1}{n_e},
\end{equation}

where, $n_e$ is the number of different loops visiting an edge $e$. For a completely packed loop (CPL) configuration (without overlap), the statistical weight of each loop is thus  $w=1$.

\subsubsection{Measurements}

\label{measurements}

The radius of gyration $R_g$ of a loop measures the typical distance between a point on a loop and its center of mass:
$$
\displaystyle R_g^2 = \frac{1}{L} \left( \sum_e {\bf X_e} - \frac{1}{L} \sum_e {\bf X_e} \right)^2.
$$

Here, $L$ is the length of the loop, and ${\bf X_e}$ is the position of the site $e$. For a self-similar loop, the radius of gyration follows a power law of the form $R_g \sim L^\nu$, where the exponent $\nu$ is related to the fractal dimension $d_f$ by the relation $\nu = 1/d_f$. The two other observables we report in the main text are the probability $C(r)$ that two sites at a distance $r$ are visited by the same loop, and the loop length distribution $P(L)$. For self-similar loops, these quantities decay algebraically as $C(r) \sim r^{-2x_2}$ and $P(L) \sim L^{-\tau}$. 

\section{Theory: computing the geometry of the Lagrangian  trajectories}

In this sections, we  recall some theoretical results and 
show how to combine them to quantitatively characterize the fractal geometry of our Lagrangian trajectories.
In subsection~\ref{6V_CPL}, we detail the exact correspondence between the six-vertex model and the Completely Packed Loop model (CPL). 
In subsection~\ref{CPL_Potts_FK}, we elaborate on the correspondences between the CPL model, the $q=1$ state Potts model, and bond percolation on a square lattice.
We finally gather  the results required to compute the main geometrical  features of our active-hydraulic flows.

\subsection{From the six-vertex model to a loop model: The Baxter-Kelland-Wu mapping}

\label{6V_CPL}

\subsubsection{Partition function of the six-vertex model}

\begin{figure}[!h]
    \includegraphics[width=\columnwidth]{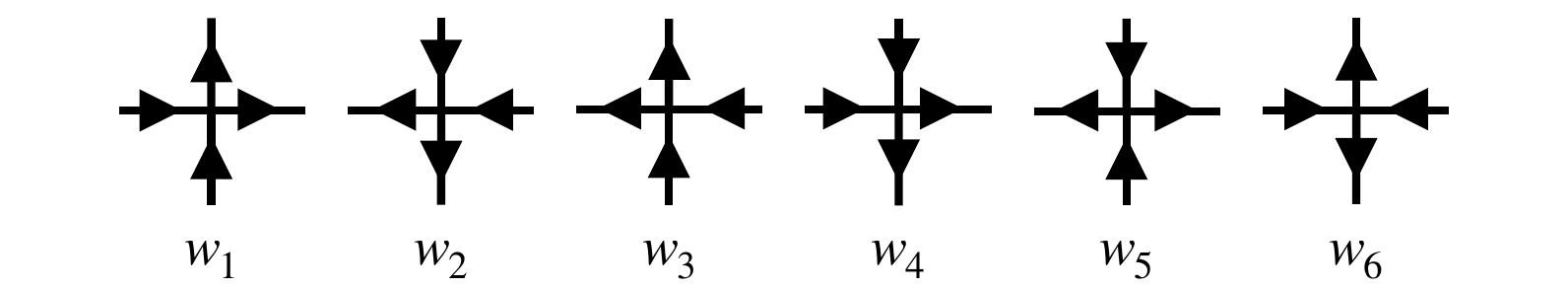}
    \caption{ {\bf The six vertex model.} The six vertices that obey the two-in-two out "ice rule" have  statistical weights $w_i$.}
    \label{6V}
\end{figure}

We recall that the six-vertex model consists in orienting the edges of a square network while satisfying the two-in-two-out "ice rule" of Figure~\ref{6V}.
Each vertex  must hence be the intersection of  two incoming and two outgoing arrows, leading to only six possible configurations. 
We then assign an energy $\epsilon_i$, $i = 1,...,6$ to each vertex. 
The total energy of the system in a configuration $\mathcal{C}$ is then given by:
\begin{equation}
    \displaystyle E(\mathcal{C}) = \sum_{i=1}^6 N_i(\mathcal{C}) \epsilon_i,
\end{equation}
where $N_i(\mathcal{C})$ is the number of vertices of type $i$ in the configuration $\mathcal C$. 
Finally, noting $w_i = \exp{(-\epsilon_i/k_BT)}$  the Boltman's weight of the $i$ vertex, the partition function of the six-vertex model takes the standard form:
\begin{equation}
    \displaystyle Z_{6V} = \sum_\mathcal{C} \prod_{i=1}^6w_i^{N_i(\mathcal{C})}
    \label{eq:Z6V}
\end{equation}
The symmetry under the  reversal of all the edge orientations imposes $w_1 = w_2$, $w_3 = w_4$, and $w_5 = w_6$, in addition, for a given set of boundary conditions one can show that $N_5(\mathcal C)-N_6(\mathcal C)={\rm cst}$ for all configurations $\mathcal C$~\cite{zinn2009six}. 
We show below that the partition function of the six-vertex model maps on the partition function of a so-called loop model.

\subsubsection{Partition function of the completely packed looped model}

\begin{figure}[!h]
\includegraphics[width=\columnwidth]{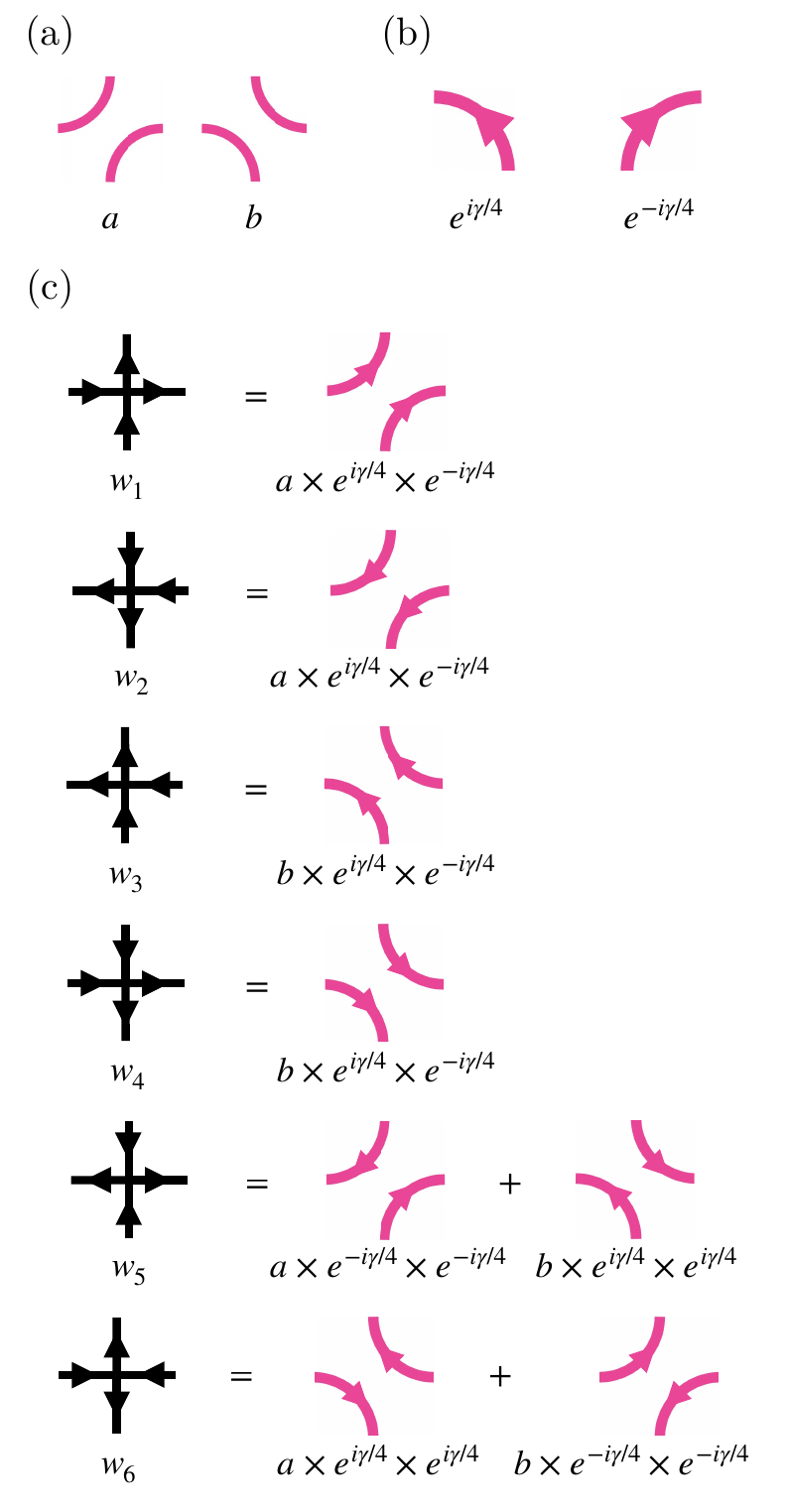}
    \caption{ {\bf Mapping between the six-vertex model and the CPL model.}
    {\bf (a)} The two vertices of the Completely Packed Loop model, and their statistical weights. 
    {\bf (b)} Statistical weights of the two elementary oriented paths that form the loops.
    {\bf (c)} Correspondance between the vertices of the six-vertex and of the CPL models.
    }
    \label{loop}
\end{figure}
Th Completely Packed Loop  model (CPL model) consists in covering all the sites of a square lattice with an ensemble of self-avoiding loops.
The partition function of the CPL model is given by:
\begin{equation}
    \displaystyle Z_{CPL} = \sum_\mathcal{C'} a^{n_1}b^{n_2} N^m,
    \label{Z:CPL}
\end{equation}
where $a$ and $b$ are the statistical weights of the two types of elementary units sketched in Fig.~\ref{loop}.
$n_1$ and  $n_2$ are the numbers of nodes with weights $a$ and $b$, and 
$m$ is the number of loops  that define  the configuration $\mathcal{C'}$. 
The statistical weight $N$ is commonly referred to as the loop fugacity. 
It is  worth noting that the statistical weight of a configuration is non-local. In other words, it cannot be computed by multiplying the statistical weight of each edge, vertex, or site of the lattice~\cite{zinn2009six}. 
\footnote{In all that follows we consider infinite square lattices and therefore avoid the subtleties  associated to the  non-contractible loops associated to periodic boundary conditions~\cite{zinn2009six}.}

\subsubsection{Mapping the six-vertex model on a Closely Packed loop model}

We briefly recall how  the six-vertex model  and the CPL model can be mapped on one another. For mode detailed discussions see e.g.  \cite{baxter1976equivalence,lafay2022geometrical,zinn2009six}. Our starting point is the partition function of the six-vertex model Eq.~\eqref{eq:Z6V}, we can express it in a local form as
\begin{equation}
Z_{6V}=\sum_{\mathcal C}\prod_{v}w(v),
\label{eq:Z6Vlocal}
\end{equation}
where the $v$s label the index of vertices, and $w(v)$ is their statistical weight ($w(v)=w_1,\,\ldots,w_6$).    
We can now use the Baxter-Kelland-Wu trick~\cite{baxter1976equivalence}. 
We associate each vertex to a pair of elementary paths illustrated in Fig.~\ref{loop}b.
The key observation is that using the mapping of Fig.~\ref{loop}b, the vertex configurations compatible with the six-vertex rules define paths that form a soup of completely packed loops. 

To make the mapping on CPL models quantitative,  we then assign to each elementary path  a weight $\exp(\pm i\gamma/4)$, where $\gamma$ is a parameter, and where the sign factor defines the  handedness of the turn. We can therefore express the statistical weights of the six-vertex model as
\begin{equation}
\begin{aligned}
    w_1 &= a  e^{i\frac \gamma 4}  e^{-i\frac \gamma 4}=a, \\
    w_2 &= a  e^{i\frac \gamma 4}  e^{-i\frac \gamma 4}=a, \\
    w_3 &= b e^{i\frac \gamma 4} \times e^{-i\frac \gamma 4}=b, \\
    w_4 &= b e^{i\frac \gamma 4}  e^{-i\frac \gamma 4}=b, \\
    w_5 &= a e^{-i\frac \gamma 4}  e^{-i\frac \gamma 4} + b e^{i\frac \gamma 4} e^{i\frac \gamma 4} = a e^{-i\frac \gamma 2} + b e^{i\frac \gamma 2}, \\
    w_6 &= a  e^{i\frac \gamma 4}  e^{i\frac \gamma 4} + b  e^{-i\frac \gamma 4}  e^{-i\frac \gamma 4} = a e^{i\frac \gamma 2} + b e^{-i\frac \gamma 2}.
    \label{eq:wiab}
\end{aligned}
\end{equation}
We can now use the expressions of Eq.~\eqref{eq:wiab} to compute the partition function of Eq.~\eqref{eq:Z6Vlocal}. 
To recast $Z_{6V}$ into $Z_{CPL}$, we use the fact that along a loop, the products of the $\exp(\pm i\gamma/4)$ reduces to $\exp(\pm i\gamma)$ and defines its global winding direction. By rearranging the product over the $v$ index to follow the loops, and noting that the two winding directions have the same probability,  we find
\begin{equation}
Z_{6V}=\sum_{\mathcal C'}a^{n_1}b^{n_2}(2\cos \gamma)^m,
\end{equation}
which exactly corresponds to Eq.~\eqref{Z:CPL} when $N=2\cos\gamma$.

\subsection{Loop models, Potts model and FK percolation}
\label{CPL_Potts_FK}

In this section we recall another series of useful mappings between the Potts model, Bond percolation and loop models.

\begin{figure}[h!]
\includegraphics[width=\columnwidth]{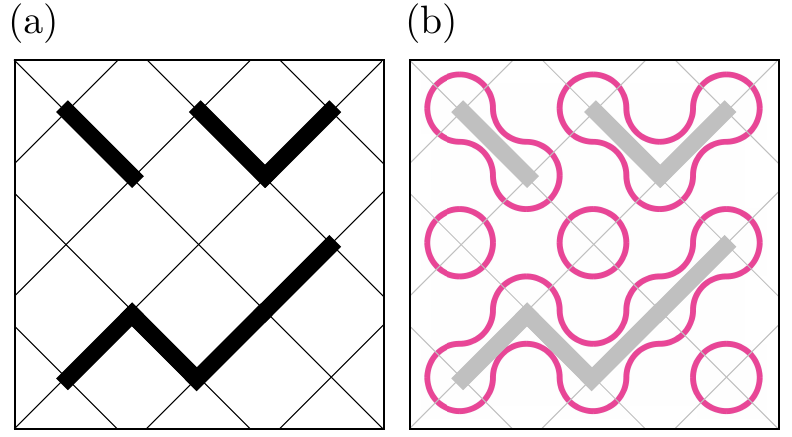}
     \caption{ {\bf Mapping the Potts model to bond percolation and CPL models} {\bf a.} Graphical representation of a Potts model  on a (tilted) square lattice. 
     The sites are indicated by dots. The thick lines represent the bonds that connect  identical spins. 
      In the limit $q \to 1$, the Potts model maps on a bond percolation model. 
     {\bf b.} The bond percolation model maps on a closely packed loop model on the  square lattice formed by the centers of the edges. By orienting the loops and using the Baxter-Kelland-Wu mapping, we recover a six-vertex model. 
     On this image, the percolation graph includes $6$ connected components (including three trivial cluster) and $7$ bonds. They result in $6$ loops on this lattice formed of $13$ sites.}
    \label{potts}
\end{figure}

\subsubsection{Partition function of the Potts model.} 
The $q$-state Potts model is a spin model where each spin $s_i$ takes a value in ${1,2,...,q}$. The energy of the system takes the form
\begin{equation}
    \displaystyle E = -J\sum_{\langle i,j \rangle} \delta_{s_i s_j},
\end{equation}
where ${\langle i,j \rangle}$ denotes the sum over nearest neighbors, and $\delta_{s_i s_j}$ is the Kronecker symbol. When $q=2$ the Potts model is equivalent to the Ising model. In general, its partition function  reads

\begin{equation}
\begin{aligned}
    \displaystyle Z_{\rm Potts} &= \sum_{\{ s \}} e^{J\sum_{\langle i,j \rangle} \delta_{s_i s_j}}
    \\
    &= \sum_{\{ s \}} \prod_{\langle ij \rangle} \left[ 1 + \left(e^J-1\right)\delta_{s_i s_j}\right]   \\
    &=\sum_{\lbrace s\rbrace} 1+(e^J-1)\delta_{s_1s_2}+(e^J-1)\delta_{s_2s_3}+\ldots   \\
    &+(e^J-1)^2\delta_{s_1s_2}\delta_{s_2s_3} +(e^J-1)^2\delta_{s_2s_3}\delta_{s_3s_4}+\ldots \\
    &+(e^J-1)^3\delta_{s_1s_2}\delta_{s_2s_3}\delta_{s_3s_4}+\ldots 
    \label{eq:Zpotts}
\end{aligned}
\end{equation}

Using the latter expansion, it was shown that 
$ Z_{\rm Potts}$ can be computed graphically using a graph representation \cite{baxter1976equivalence,nienhuis1987coulomb,henkel2012conformal,lafay2022geometrical}.
To do so, we draw lines along the edges that connect spins of identical values, see Fig.~\ref{potts}a. Each term in the sum of Eq.~\eqref{eq:Zpotts} defines a subgraph $G$ on the square lattice. 
We  therefore note that 
the sum over the spin configurations can be written as a weighted sum over all the possible subgraphs
\begin{equation}
    Z_{\rm Pott}=\sum_{ \{ G \} }(e^J-1)^e q^c ,
    \label{eq:Zwhithney}
\end{equation}
where $e$ is the number of bonds in the graph and $c$ the number of connected components.
For example, in the configuration shown in Fig.~\ref{potts}b  $c=6$  and $e=7$.  Eq.~\eqref{eq:Zwhithney} is the so called Whitney polynomial expansion of the Potts model partition function.

\subsubsection{Correspondence with bond percolation.} 
The above representation of the Potts model leads to a correspondence with the bond-percolation problem.
In the limit $ q \to 1$, Eq.~\eqref{eq:Zwhithney} is indeed  the partition function of the bond percolation on a the square lattice, with the bond occupation probability being $(e^J-1)/e^J$~\cite{baxter1976equivalence}.

\subsubsection{Mapping the Potts model on the CPL model} 
Finally, we show that the partition function of the Potts model is equal to that of the CPL model \cite{nienhuis1987coulomb,henkel2012conformal,lafay2022geometrical}. 
In Fig.~\ref{potts}b, we show how to dress a graph $G$ with closely packed loops. We can then use the Euler formula to relate  the number of loops ($m$), the total number of sites($n_{\rm s}$), the number of connected components of the graph ($c$) and the number of bonds  ($e$): $m = 2c + e - n_{\rm s} $ 
The graph showed in Fig.~\ref{potts}a contains $6$ clusters, $7$ bonds, $13$ sites. The Euler formula correctly predicts a number of 6 loops.
We can then recast the partition function of the Potts model (Eq.~\eqref{eq:Zwhithney})  as:

\begin{equation}
\begin{aligned}
     Z_{\rm Potts} &= \sum_{{\mathcal{C}_{\rm P}}} (e^J-1)^e q^c \\
     &= \sqrt{q}^{n_{\rm s}} \sum_{{\mathcal{C}_{\rm P}}} \left( \frac{e^J-1}{\sqrt{q}}\right)^e\sqrt{q}^m \\
      &= \sqrt{q}^{n_{\rm s}} Z_{CPL}\left(\frac{a}{b}=  \frac{e^J-1}{\sqrt{q}}, N=\sqrt{q}\right),
\end{aligned}
\end{equation}
thereby demonstrating the exact mapping between the Potts and CPL models.


\subsection{Geometry of the Lagrangian trajectories}

\label{geometrie_CPL}
We can now take advantage of the series of mappings recalled above to predict the geometry of the Lagrangian trajectories of our active hydraulic flows. 
To achieve this goal we use the non-perturbative results of 
Saleur and Duplantier who have calculated exactly the fractal dimension of the Hull of the percolation clusters in dimension $2$~\cite{saleur1987exact}.
They used the series of analogies described above: at the critical point, the perimeter of the hull of the largest  percolation cluster  scales like the length of Baxter-Kelland-Wu loops. 
Combining this observation to technical Coulomb-gas calculations, they found that their fractal dimension should be $D_H = 7/4$. 

Independently, Aharony and coworkers showed numerically~\cite{grossman1986structure}, and   analytically~ \cite{aizenman1999path} that the "external" perimeter of the critical percolation clusters has a fractal dimension $D_e = 4/3$. Kolb and Rosso then showed that any loop  inside a percolation cluster has a fractal dimension equal to $D_e = 4/3$ \cite{kolb1993loop}. 

Equipped with these formal results, we can now predict the geometry of the flows measured in our experiments.
Firstly, the largest trajectories of of hydraulic flows correspond to completely packed loops. Their hull should therefore have a fractal dimension $D_H = 7/4$, and thus a $\nu$ exponent $\nu = D_H^{-1} = 4/7 \approx 0.57$ in excellent agreement with our measurements. 
Secondly, the smaller loops should have a fractal dimension equal to $D_e = 4/3$, and a $\nu$ exponent $\nu = 3/4$ again in excellent agreement with our experimental measurements. 

Beyond the prediction of their gyration radius, we can predict the distribution of the length of the loops $P(L)\propto L^{-\tau}$ and the two point correlation function $C(r)\propto r^{-2x_2}$ defined in the main text. 
They exponents are determined by the hyperscaling relations $x_2 = 2 - 1/\nu = 2/3$ and $\tau - 1 = 2\nu = 5/2$  predicted in \cite{saleur1987exact,kondev1996operator}.
The good agreement with our experimental findings provide yet another confirmation of the predictive power of our active-hydraulic laws.

\bibliographystyle{apsrev4-1}
\bibliography{biblio_SI}